\documentclass[a4paper]{article}
\usepackage[dvips]{graphicx}
\usepackage{amssymb}

\begin{document}
\title{Link smearing considered as MCRG transformation}
\author{F. Geles\footnote{faruk.geles@uni-graz.at} and 
C.B. Lang\footnote{christian.lang@uni-graz.at}}
\maketitle
\begin{center}
Institut f\"ur Physik (Theoretische Physik)\\
Universit\"at Graz,\\
A-8010 Graz, Austria\\
~\\
\end{center}
\begin{abstract}Gauge link smearing is widely used in lattice QCD computations.  The
idea is to remove the local (UV) fluctuations of the gauge field configurations
while keeping the longer-range (IR) properties intact.  Important applications
are in the definitions of interpolating hadron operators  as well as in
updating the gauge field configurations with dynamical quarks in the so-called
Hybrid Monte Carlo (HMC) algorithm. Here we study the effectiveness of various
smearing methods and try to quantify these with tools known from Monte Carlo
Renormalization Group (MCRG).
\end{abstract}

\section{Motivation} 

In lattice gauge theory the group valued link variables encode the gauge field. In the Monte
Carlo (MC) approach the space of gauge configurations is sampled according to
the probability weight factor $\exp{(-S)}$, where $S$ denotes the effective
gauge action, which may include terms coming from the integration of the
fermionic variables as well. All observables are then determined from averaging
functions of the gauge link variables over the MC-generated ensemble of gauge
configurations. Wilson loops (traces of ordered products of gauge link
variables along a loop) are the best known of such observables and allow to
estimate the string tension. However, also quark propagators (leading to hadron
n-point functions) are computed from the link variables.

Since in many cases one is interested in the infrared behavior, i.e.,
correlations over distances large in lattice units, it is often worthwhile to
smoothen the gauge configuration in a gauge-invariant fashion. This is done by
constructing new link variables from an averaged sum of gauge link products
along paths connecting the end points of the original link. Since the lattice size is not
changed by this procedure we can consider such a smearing step as a MC block spin transformation
(BST) with scaling factor 1. Actually, if we would be able to determine the
effective action corresponding to the smeared configurations weight factor, one
could simulate directly the smoother gauge system, without changing long 
distance correlations.

Smearing involves neighboring gauge links and thus has a certain effective
radius of impact in lattice units. As long as the number of smearing steps is
small compared to the lattice size, the long distance  correlations will be
unaffected. Iterating the smearing procedure extends this range and eventually
the  configurations will approach trivial ones, where all physical information
is washed out. Non-local properties like topological modes will survive more of
such steps and indeed smearing has also been used to improve identification of
topological sectors.

Here we will study the behavior of some observables under various smearing BSTs
with the tools of the Monte Carlo Renormalization Group (MCRG) and try  to
obtain a more quantitative characterization of the smearing flow and efficiency
of these smearing methods.

For completeness we summarize the four smearing algorithms that we have studied.

\paragraph{APE Smearing \cite{Albanese:1987ds}:}

The ``smeared'' link variable is built from a weighted
sum over ``staples'' and the original link variable,
\begin{eqnarray}
\tilde{U}'_{\mu}(n)&=&\mathrm{Proj}_{SU3}
\left[(1-\alpha)U_{\mu}(n)+\frac{\alpha}{6}\sum_{\mu\neq\nu}C_{\mu\nu}(n)\right]\;,\\
  C_{\mu\nu}(n)&=&U_{\nu}(n)U_{\mu}(n+\hat{\nu})U^{\dag}_{\nu}(n+\hat{\mu})\nonumber\\
  &+& U^{\dag}_{\nu}(n-\hat{\nu})U_{\mu}(n-\hat{\nu})U_{\nu}(n-\hat{\nu}+\hat{\mu})\;.
\end{eqnarray}
This sum is projected into $X\in$SU(3) by maximizing 
$\mathrm{Re}\ \mathrm{tr}[X\tilde{U}'_{\mu}(n)^{\dag}]$.
We take $\alpha=0.55$ as discussed in \cite{Bonnet:2001rc}.

\paragraph{Hypercubic (HYP) Smearing \cite{Hasenfratz:2001hp}:}

This is a 3-step procedure which takes into account contributions from within a
hypercube, again projected to SU(3). For the parameters involved in the
definition we choose the values $\alpha_1=0.75$, $\alpha_2=0.6$,
$\alpha_2=0.3$,  suggested in \cite{Hasenfratz:2001hp}.

\paragraph{nHYP Smearing \cite{Hasenfratz:2007rf}:}

This is like HYP smearing, with a projection into U(3) via
\begin{equation}
\mathrm{Proj}_{U(3)}[U']=U'\frac{1}{\sqrt{U'^{\dag}U'}}\;.
\end{equation}
This smearing is
differentiable with regard to the link variables, a property which is required
for the HMC algorithm. The parameters are chosen like for the HYP smearing.

\paragraph{STOUT Smearing \cite{Morningstar:2003gk}:}

This also leads to a differentiable expression. The new link variable
\begin{equation}
U^{'}_{\mu}(n)=\exp(iQ_{\mu}(x))U_{\mu}(n)
\end{equation}
is constructed from the hermitian, traceless matrix $Q_\mu$ which is built from
a weighted sum over staples $C_\mu(n)$. This type of smearing is also widely
used in simulations with the HMC algorithm. Following the notation  of
\cite{Morningstar:2003gk} we choose $\rho=0.1$ as suggested in
\cite{Durr:2004xu}.

\section{MCRG tools}
 
We study the different methods on an ensemble of 300 configurations of lattice
size $16^3\times 32$, generated with two light dynamical quarks flavors and
the Chirally  Improved Dirac operator
\cite{Gattringer:2000js,Gattringer:2000qu}. The parameters of the action
correspond to a lattice unit of 0.144 fm and a pion mass of 322 MeV. Details on
action and simulation can be found in \cite{Gattringer:2008vj,Engel:2010my}.

The expectation values of five observables $S_i$ are determined: plaquette $S_1$,
planar rectangular $2 \times 1$-loop $S_2$, bent loop $S_3$, twisted bent loop
(``chair'') $S_4$ and the topological charge  $S_5$ and susceptibility (in the
formulation of \cite{Bonnet:2001rc}).

\begin{figure}[t]
\begin{center}
\includegraphics*[width=0.45\textwidth,clip]{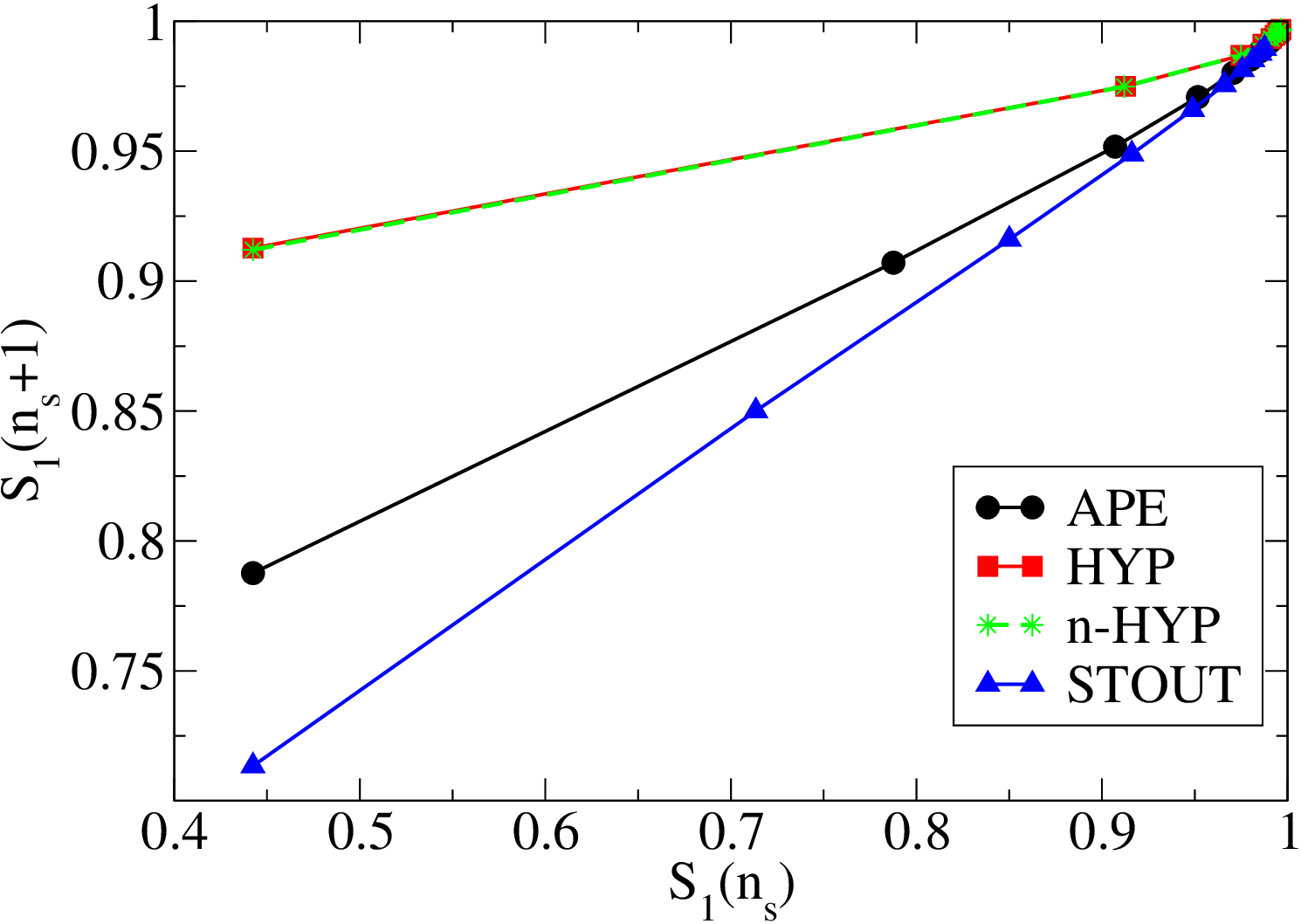}
\hfill
\includegraphics[width=0.45\textwidth,clip]{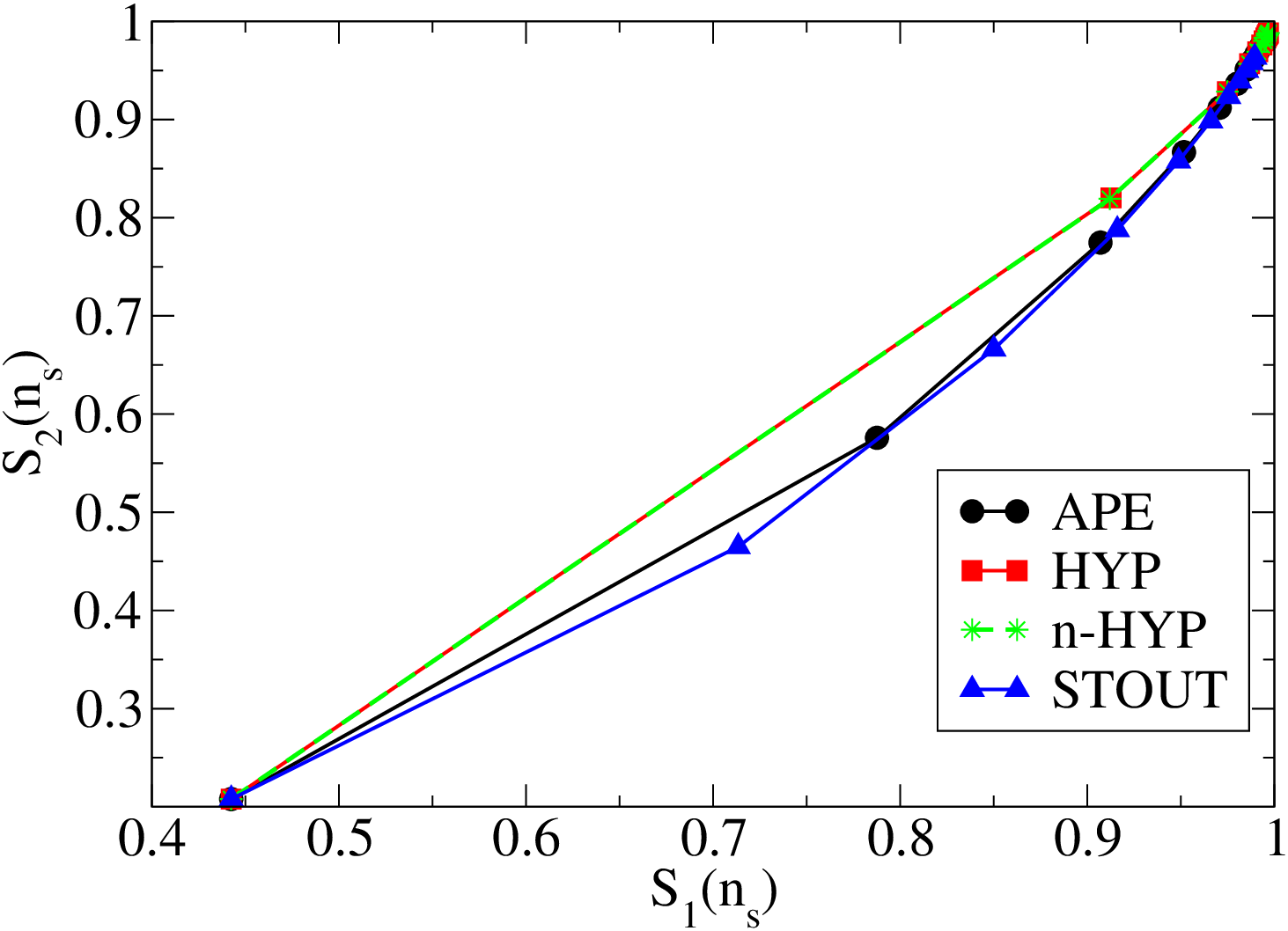}
\end{center}
\caption{\label{fig_step}
L.h.s.: The ``step smearing function'' for the plaquette observable $S_1$,
showing the mapping $S_1(n_s)\mapsto S_1(n_s+1)$, where $n_s$ denotes the number
of smearing step (left-most point corresponds to $n_s=0$). The points are
connected by straight lines to guide the eye. R.h.s.: The subsector of $S_1$ vs.
$S_2$ in the space of observables is shown to follow trajectories that are
close to each other for the four smearing types considered.}
\end{figure}

Figure \ref{fig_step} (l.h.s.) shows the change of the plaquette  expectation
value from one step of smearing to the next (for all four methods).  Already
here one finds a ratio  of HYP\,:\,APE\,:\,STOUT $\approx$ 
3\,:\,2\,:\,1 (i.e., one HYP step corresponds approximately to three
STOUT steps). The other observables show similar behavior.

We note that for given configurations different smearing methods lead to
{\em different} (close to integer) values of the topological charge for that
configuration, stabilizing at a value typically only after 50--100 smearing steps. This
nonlocal quantity thus proved not very useful for the subsequently discussed
flow study.

In MCRG block spin transformations \cite{Ma76} the RG flow in the space of
couplings approaches a renormalized trajectory connecting fixed points. This
renormalized trajectory has a correspondence in the space of observables.
Similarly, the smearing flow can also be followed in the space of observables
(figure \ref{fig_step}, r.h.s.). These trajectories (also for other projections
in the space of observables) are close to each other for the different smearing
types, while the speed along the trajectories can be different. 
We also find that HYP and nHYP are almost indistinguishable.

In MCRG the block spin transformation of an ensemble of spin configurations
distributed according to an action leads $\sum_i\beta_i S_i$ to another ensemble
with a different effective action $\sum_i\beta_i' S_i$, thus following a
renormalized trajectory (connecting fixed points)  in the space of coupling
constants $(\beta_i)$. We use this framework to study the change of the
effective action under smearing transformations. The derivative matrix
\begin{equation}
T_{jk}^{(n+1,n)}=
\left(\frac{\partial \beta_{j}^{(n+1)}}
{\partial \beta_{k}^{(n)}}\right)
\end{equation}
(where the superscripts denote the number of blocking steps)
defines a linearization of the flow in coupling space. It
may be derived from cross-correlations of observables measured on 
configurations for different numbers of smearing steps \cite{Sw79},
\begin{eqnarray}
\frac{\partial  \left\langle S_{i}^{(n+1)}\right\rangle}{\partial \beta_{k}^{(n)}}&=&
 \sum_{j}\frac{\partial \beta_{j}^{(n+1)}}{\partial \beta_{k}^{(n)}}
\frac{\partial \left\langle S_{i}^{(n+1)}\right\rangle}{\partial \beta_{j}^{(n+1)}}\;, \\ 
\frac{\partial \left\langle S_{i}^{(n+1)}\right\rangle}{\partial \beta_{k}^{(n)}}&=&
 \left\langle S_{i}^{(n+1)}S_{k}^{(n)}\right\rangle - 
 \left\langle S_{i}^{(n+1)}\right\rangle\left\langle S_{k}^{(n)}\right\rangle\;. 
\end{eqnarray}
For BSTs close to a fixed point $T^{(n+1,n)}$ approaches the fixed point value $T^*$ and then its eigenvalues
$\lambda_i$ (if greater than 1)
can be related to the relevant critical exponents through 
$\nu_i=\ln b/\ln \lambda_i$ (here b is the scale factor of the BST).

In our case we are interested in $T^{(n_s+1,n_s)}$ for small $n_s$. Its
diagonalization gives the information about the main flow directions and the
speed of flow. Actually, in case one knows the starting point in coupling space
$(\beta_i)$ one can approximately recover $(\beta_i)'$. Related problems have
been discussed recently in the context of trivializing maps
\cite{Luscher:2009eq}. For the gauge configuration ensembles used here we do
not know $(\beta_i)$, since they result from a simulation with dynamical
fermions and thus the full effective gauge action is not the simple one used for
the gauge part alone.

\begin{figure}[t]
\begin{center}
\includegraphics[width=0.32\textwidth,clip]{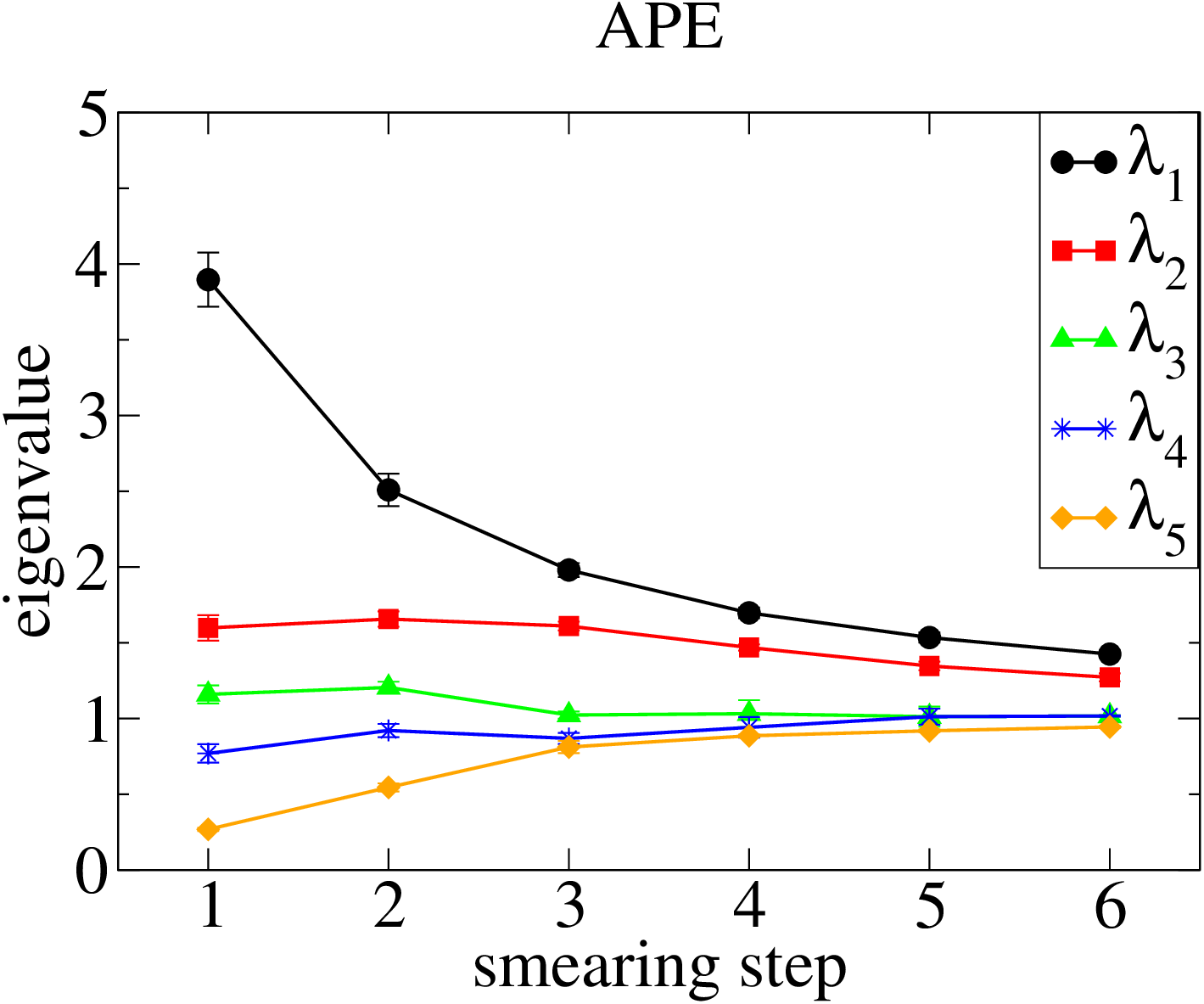}
\includegraphics[width=0.32\textwidth,clip]{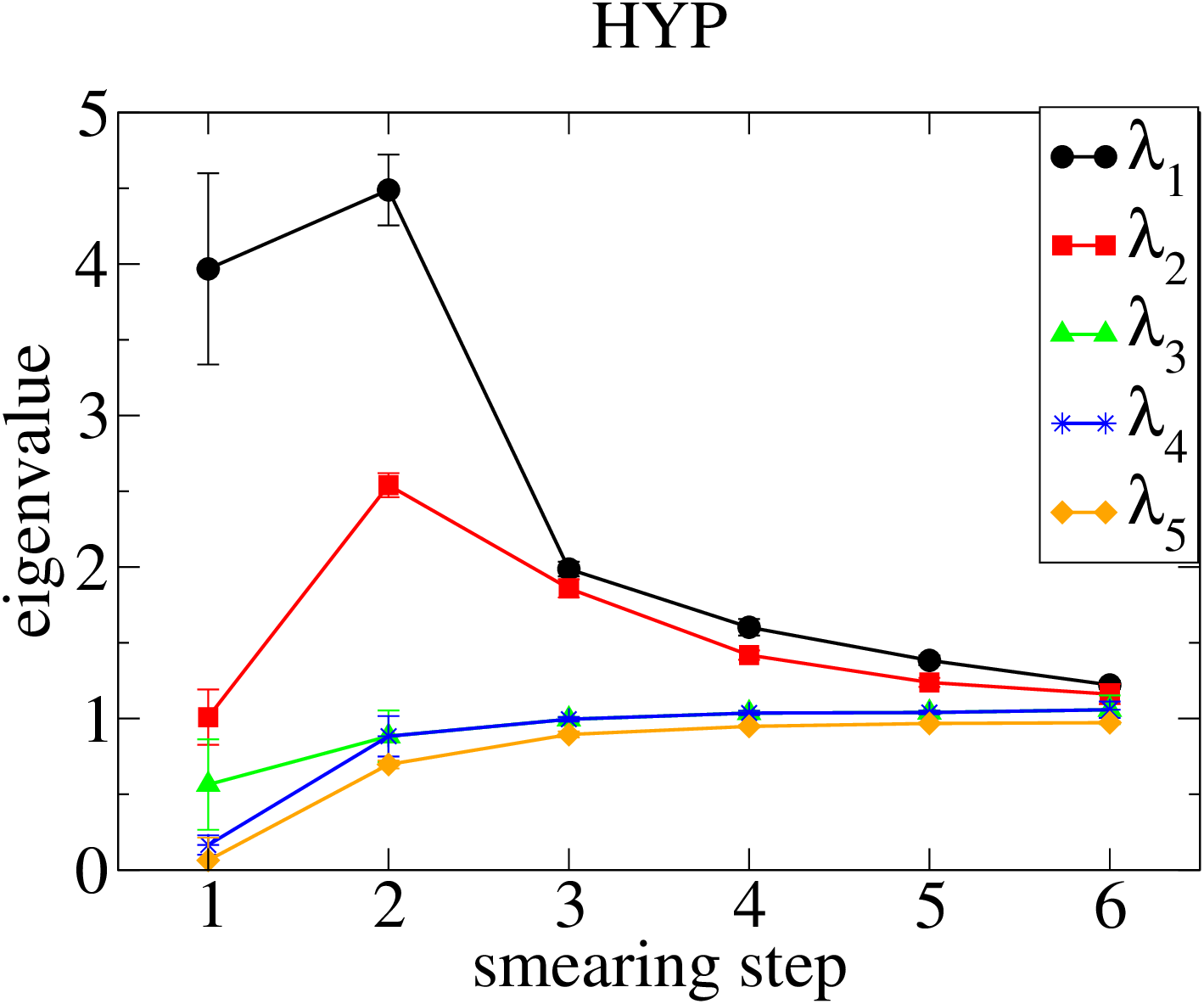}
\includegraphics[width=0.32\textwidth,clip]{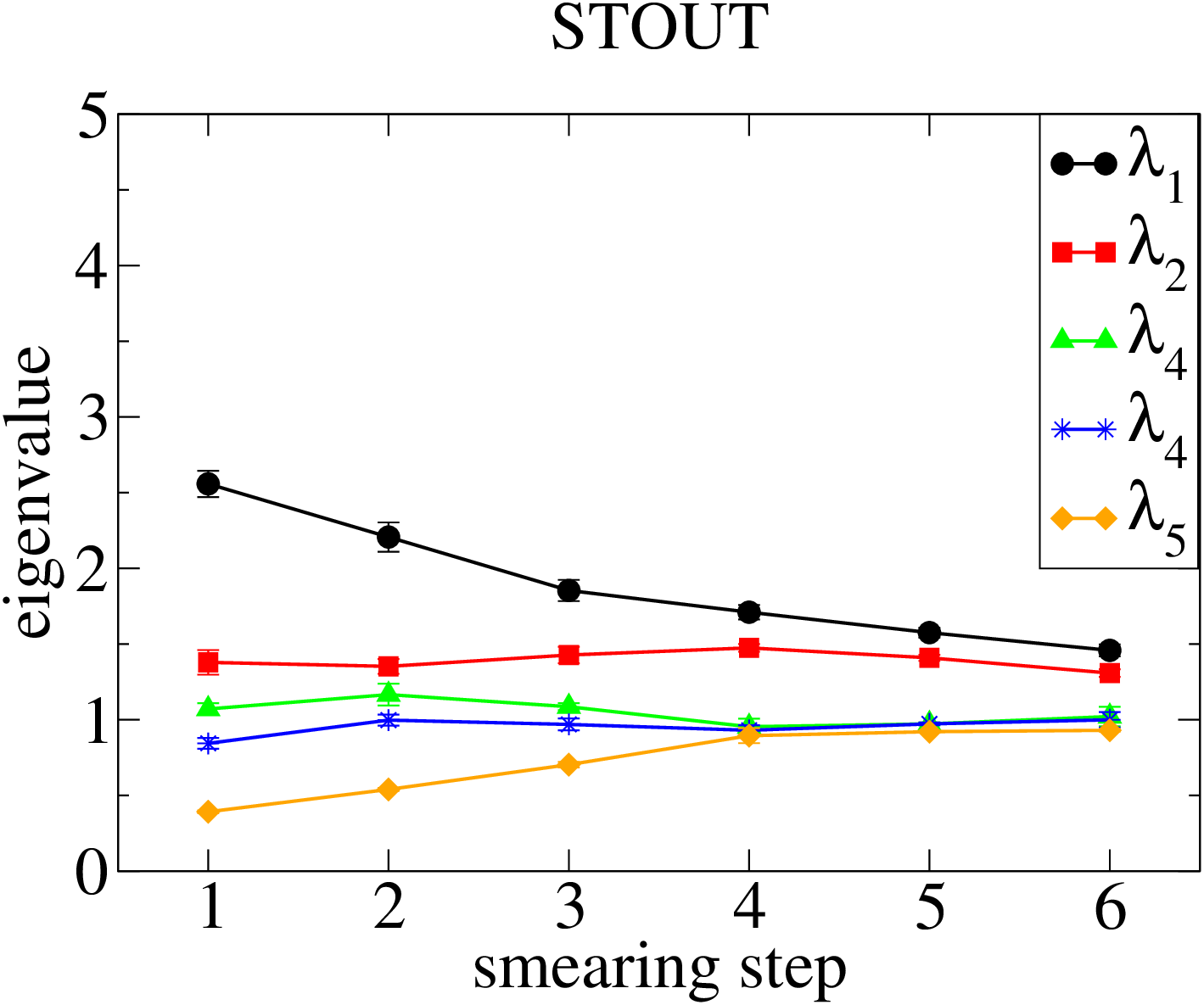}
\end{center}
\caption{\label{fig_eigval}
The figures show the eigenvalues of $T^{(n+1,n)}$ (nHYP results are close to HYP).
}
\end{figure}

\begin{figure}[t]
\begin{center}
\includegraphics[width=0.32\textwidth,clip]{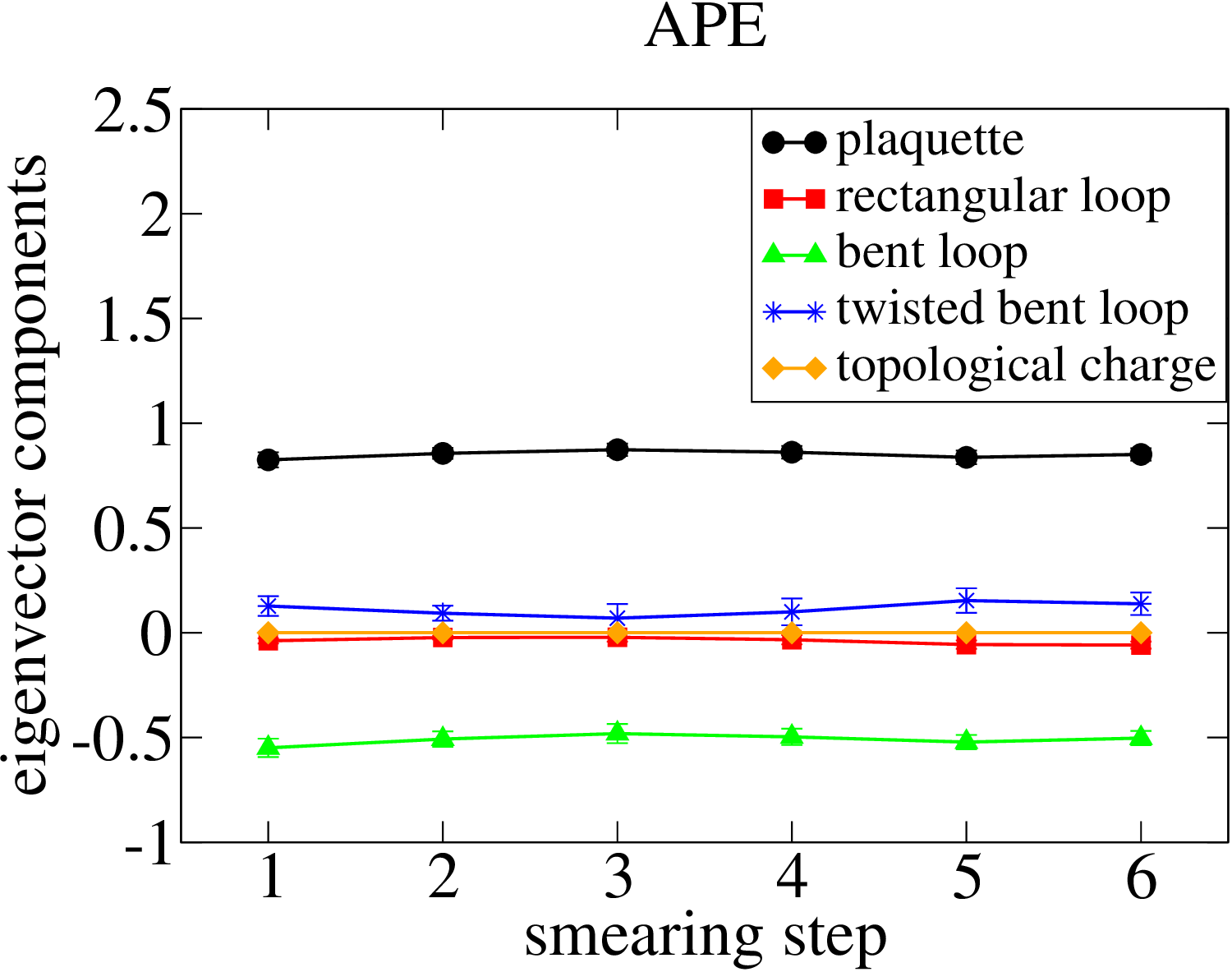}
\includegraphics[width=0.32\textwidth,clip]{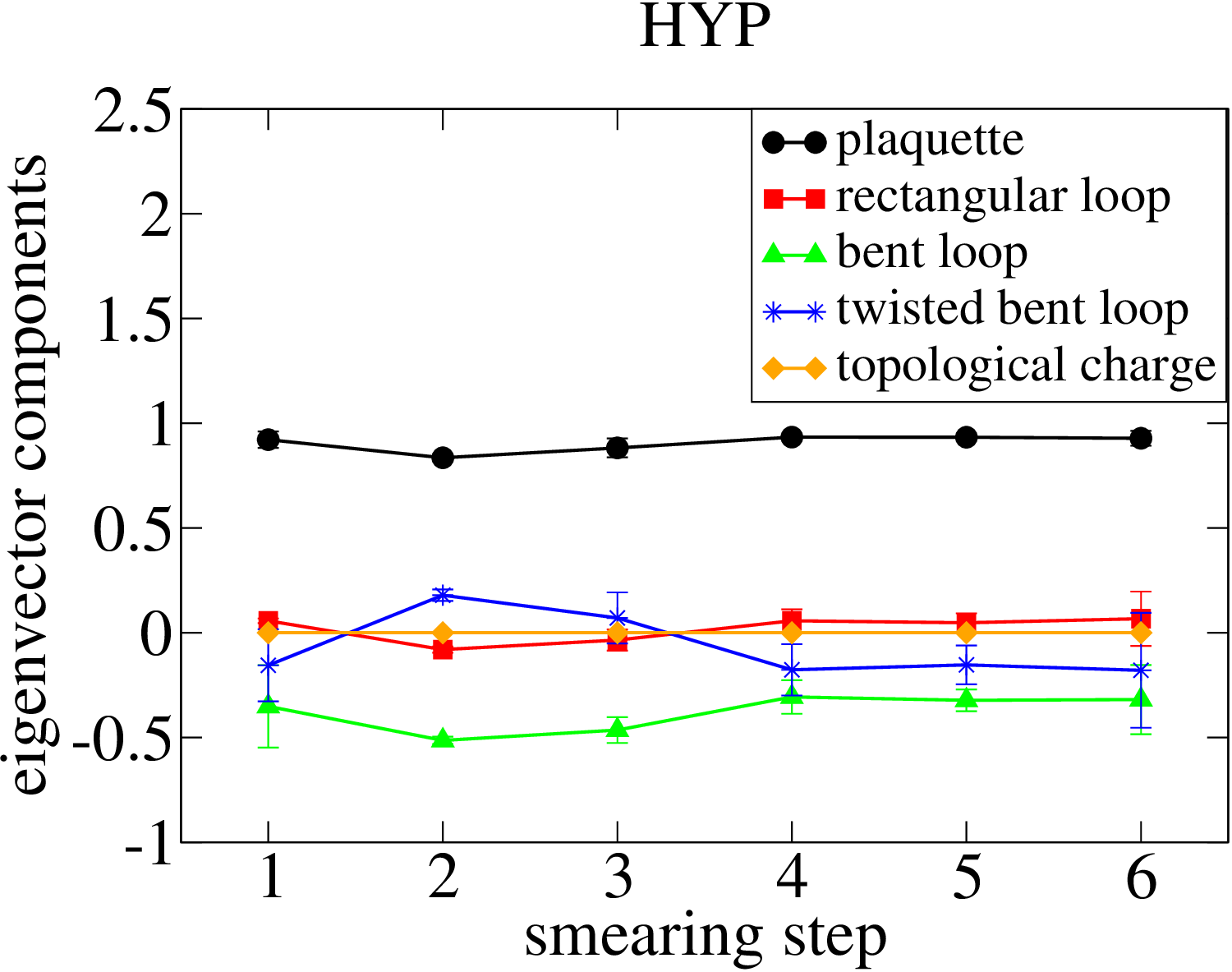}
\includegraphics[width=0.32\textwidth,clip]{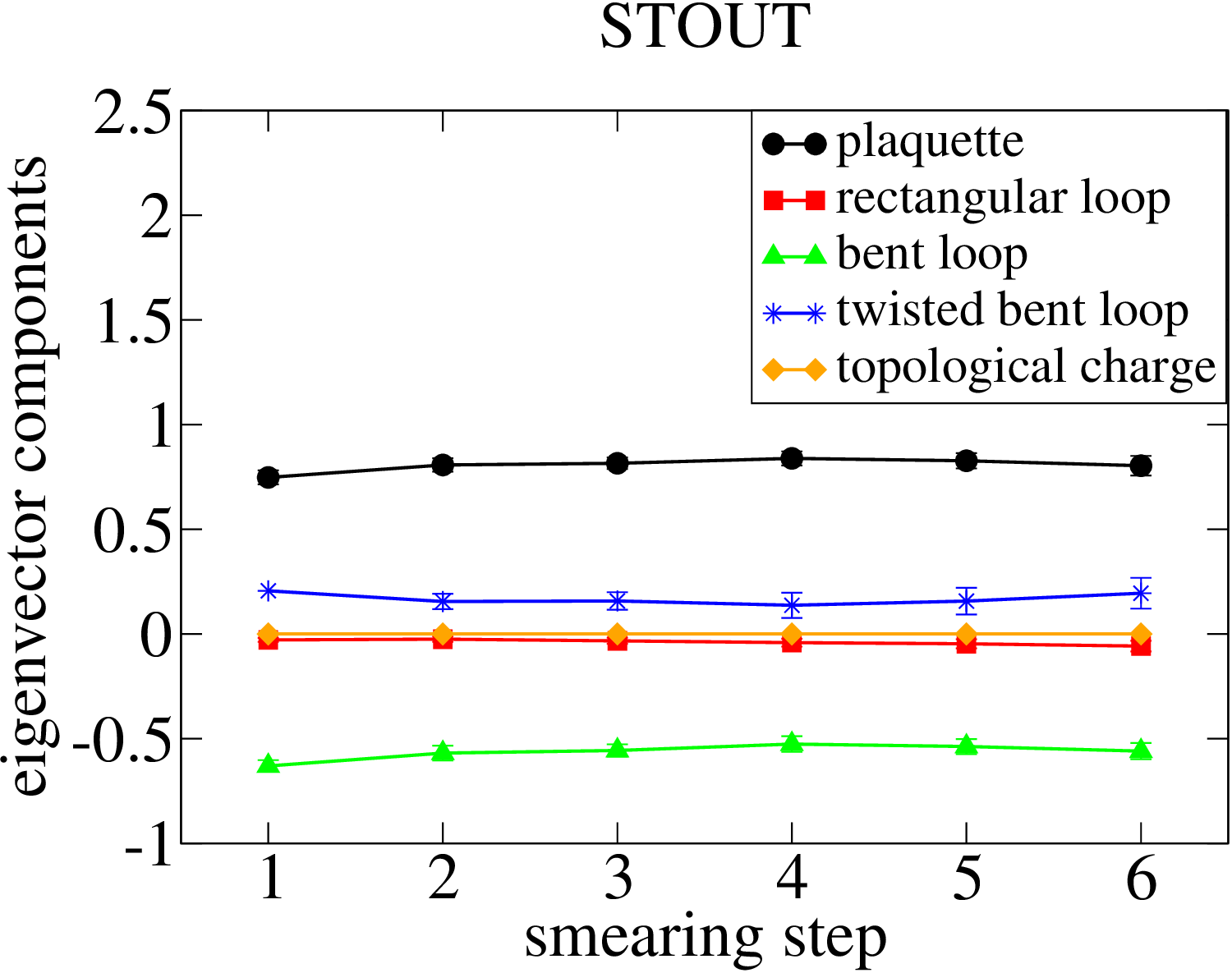}
\end{center}
\caption{\label{fig_eigvec1}
Here the eigenvector components of the leading eigenvalue of $T^{(n+1,n)}$ are plotted
(nHYP results are not shown here but are close to HYP).
}
\end{figure}

Figure \ref{fig_eigval} shows the leading eigenvalues of the diagonalization.
Only two of them are significantly larger than 1; asymptotically they all
approach 1 of course. The data for HYP smearing show a peak in the second step,
they also give the largest eigenvalues indicating the strongest effect (on the
observables considered). 

Comparing the eigenvalues with the (normalized) eigenvectors for the leading
eigenvalue (figure \ref{fig_eigvec1}) we find that the dominant direction is the
plaquette $S_1$ followed by the bent loop in all cases. The eigenvectors are
defined only up to normalization (we normalized to unit length) and overall
sign, which we choose such that continuity in $n_s$ is obtained.

The eigenvectors for the 2nd largest eigenvalue  confirms that APE and STOUT
smearing follow similar trajectories.  Here the bent loop and the twisted bent
loop dominate the flow. The eigenvectors for HYP smearing exhibit stronger
fluctuation: more directions in coupling space are activated, 
mixing the local degrees of freedom stronger than the other types.

\section{Conclusions}

We try to make a step towards a systematic classification of gauge link smearing
algorithms. Viewing smearing steps as analogues of block spin transformations,
we attempt to quantify their effectiveness. This is a first attempt towards
developing a tool for classification of the effect smearing has in configuration
space. We find that the different methods follow similar trajectories in the
space of local observables and effective couplings, however with different
speed. The nHYP results are close to the HYP
results. In our study only few local observables have been included; considering
more  would allow for a better and more complete quantification.

\paragraph{Acknowledgments:}

We wish to thank G. Engel, C. Gattringer and A. Hasenfratz for discussions. The
computations were done on the computing facilities of the computer center of
Universit\"at Graz.



\providecommand{\href}[2]{#2}\begingroup\raggedright
\endgroup
 
\end{document}